\documentclass[a4paper,10pt,twoside]{cpc-hepnp}

\usepackage{fancyhdr}            %页眉和页码
\usepackage{multicol}            %用于双栏排版
\usepackage{upgreek}             %希腊字母的输入
\usepackage{indentfirst}         %支持首行缩进
\usepackage{booktabs}            %画三线表
\usepackage{array,tabularx}      %支持各种表格
\usepackage{keyval,graphicx}     %用于插图
\usepackage{textcomp}            %各种符号的输入
\usepackage{amssymb,bm,mathrsfs,bbm,amscd} %各种数学符号的输入
\usepackage[tbtags]{amsmath}     %AMSLaTeX 数学公式的录入，[tbtags]用于公式编号与公式的最后一行对齐
\usepackage{lastpage}            %计算总页数
\usepackage[numbers,sort&compress]{natbib}  %参考文献的上标
\usepackage{color}

\begin{document}

%===================================================================================================================
\setlength{\abovecaptionskip}{4pt plus1pt minus1pt}   %图形中的图与标题之间的距离
\setlength{\belowcaptionskip}{4pt plus1pt minus1pt}   %表格中的表与标题之间的距离
\setlength{\abovedisplayskip}{6pt plus1pt minus1pt}   %公式前的距离
\setlength{\belowdisplayskip}{6pt plus1pt minus1pt}   %公式后面的距离
\addtolength{\thinmuskip}{-1mu}            %减小自变量与函数关系的间距，比如sin x
\addtolength{\medmuskip}{-2mu}             %减小+,-,\times,\div的两侧空白
\addtolength{\thickmuskip}{-2mu}           %减小等号、不等号的两侧空白
\setlength{\belowrulesep}{0pt}          %在使用booktabs宏包画三线表中，加了 | 来加竖线时，会使横线和竖线不能很好
\setlength{\aboverulesep}{0pt}          %的交叉，使用这两个设置即可。
\setlength{\arraycolsep}{2pt}           %减小eqnarray中＝号两边的距离

\providecommand{\e}[1]{\ensuremath{\times 10^{#1}}}

%===================================================================================================================

%\fancyhead[c]{\small Chinese Physics C~~~Vol. 37, No. 3 (2013) 035101}
%\fancyfoot[C]{\small 035101-\thepage}

\footnotetext[0]{Received 9 April 2012, Revised 16 September 2012}

\title{\boldmath Legendre transformations and {the} thermodynamic geometry\\ of 5D black holes\thanks{Supported by
Scientific and Technological Foundation of Chongqing Municipal Education Commission (KJ100706)}}

\author{%
HAN Yi-Wen$^{1)}$\email{hanyw1965\oa 163.com}
\quad CHEN Gang
\quad LAN Ming-Jian%
}

\maketitle

\address{%
College of Computer Science, Chongqing Technology and Business University, Chongqing 400067, China\\
}

\begin{abstract}
This paper studies the thermodynamic properties of the 5D
black hole in Einstein-Gauss-Bonnet gravity from the viewpoint of
geometrothermodynamics. It {is found} that the Legendre invariant metrics of the
5D black {holes} in Einstein-Yang-Mills-Gauss-Bonnet {theory and }
Einstein-Maxwell-Gauss-Bonnet {theory} reproduce the behavior of the
thermodynamic interaction and phase transition structure of the
corresponding black hole configurations {correctly}. It is shown that they are both
curved and {that} the curvature scalar {provides} information about the phase
transition point.
\end{abstract}


\begin{pacs}
04.70.Dy, 04.20.-q, 04.50.+h \qquad {\bf DOI:} 10.1088/1674-1137/37/3/035101
\end{pacs}

\footnotetext[0]{\hspace*{-3mm}\raisebox{0.3ex}{$\scriptstyle\copyright$}2013
Chinese Physical Society and the Institute of High Energy Physics
of the Chinese Academy of Sciences and the Institute
of Modern Physics of the Chinese Academy of Sciences and IOP Publishing Ltd}%

\begin{multicols}{2}

\section{Introduction}

{In} 1975, Weinhold proposed a geometrical way of studying thermodynamics and
gave a Riemannian metric, defined as the second derivatives of internal
energy [1]. According {to} Weinhold's theory, an interesting inner product of the
equilibrium thermodynamic state space in the energy representation was
{the} Hessian matrix of internal energy $U$ with
respect to the extensive thermodynamic variables $N^a$, namely
$g_{ij}^W=\vpartial\nolimits_i \vpartial\nolimits_j M(U,N^a)$. However, there was no physical
interpretation associated with this metric structure [2]. As a modification,
Ruppeiner introduced a Riemannian metric ($g_{ij}^R =-\vpartial\nolimits_i \vpartial\nolimits_j
S(U,N^a))$ into {the} thermodynamic system once more, and defended it as the second
derivative of entropy $S$ (here, the entropy is a function of the internal
energy $U$ and other extensive variables $N^a$ of a thermodynamic system) [3].
Based on {Ruppeiner's} theory above, and including {the} Weinhold metric,
Ferrara et al. [4] investigated the critical points of moduli space by using
{the} Weinhold and Ruppeiner {metrics}. Since then, {black} hole
thermodynamic geometry has been one of the focuses in theoretical physics.
Until now, there {have been many} geometrical descriptions of equilibrium
thermodynamics. For example, Aman et al. [5--7]
showed the relation between {the} thermodynamic (Riemannian) curvature of the
Reissner-Nordstrom black hole and the higher dimensional black hole. Sarkar
et al. [8, 9] gave a brief review on the geometrical method {of}
thermodynamics, and applied this approach to the BTZ black hole and extremal
black holes in string theory. Their studies showed that Ruppeiner geometry
can overcome problems such as {the} covariance and self-consistency of general
thermodynamics, {which} has {the} phase structure information of {the}
thermodynamic system, and {this} was applied into all kinds of thermodynamic
modes [10--13]. In addition, Ruppeiner {gave} a systematic discussion on
how to make the correct metric {choice}, {and also} demonstrated
several limiting results matching extreme Kerr-Newman black hole
thermodynamics to the {two-dimensional} Fermi gas, showing that connection to
a 2D model is consistent with the membrane paradigm of black holes [14, 15].
With Ruppeiner's thermodynamics geometry theory, it is shown that Ruppeiner
geometry can {perform} the physical meanings of various thermodynamic
systems [16--22], such as the ideal gas, the van der Waals gas, and so on. The
results revealed {the} fact that the scalar curvature is zero and the Ruppeiner
metric is flat, for the van der Waals gas, while the curvature is non-zero
and diverges only as the phase transition take place. The focus of the above
problems is on the thermodynamic potential, which is generally believed to
be the internal energy rather than the simple {mass}. But {the} above studies
have shown that {Weinhold} and Ruppeiner's thermodynamic metrics are not
invariant under Legendre transformations.

Recently, Quevedo et al. [23] presented a new formalism of
geometrothermodynamics (GTD) as a geometric approach that incorporates
Legendre invariance in a natural way, and allows us to derive Legendre
invariant me-

\end{multicols}
\begin{multicols}{2}

\noindent trics in the space of equilibrium states. Considering the
Legendre invariant, they presented a unified geometry
where the metric structure can give a {good} description
of various types of black hole thermodynamics [24--26]. {One of the aims} of
the application of different thermodynamic geometries is to
describe phase transitions in terms of curvature singularities.
For a thermodynamic system, it is quite interesting to investigate
the corresponding relationship between the curvature of {the} Weinhold
metric, {the} Ruppeiner metric and the Legendre invariant metric, and the
phase transitions. In fact the above viewpoints have been applied
to various black holes [27--32], {even} though it is widely
believed that the thermodynamic geometry of a black hole is still
a most fascinating and unresolved subject at present. The main
purpose of the present work is to show that the Legendre invariant
metric can be used to correctly {reproduce} the thermodynamics of
the 5D black hole and the EMGB black hole. This case has been
analyzed previously {by} using a different approach in which
Legendre invariance is not taken into account [22].

The organization of this paper is outlined as follows. We show {the} GTD of the 5D
black hole in Einstein-Yang-Mills-Gauss-Bonnet (EYMGB) theory in Section~2.
Then in Section~3, the GTD of the 5D black hole in Einstein-Maxwell-Gauss-Bonnet
(EMGB) {theory is} described. Section~4 {provides} some discussions and conclusions.
Throughout this paper, the units $c=\hbar =G=1$ are used, {and} the computer
algebra system Mathematica 7.0 was used for most of the {calculations}.

\section{Geometrothermodynamics of the 5D black hole in EYMGB theory}

In this section we first describe the black hole solution in
Einstein-Yang-Mills-Gauss-Bonnet (EYMGB) theory and then study the
thermodynamic properties {in} the next subsection. In EYMGB gravity theory, the
5D spherically symmetric solution obtained recently by Mazharimousavi and
Halisoy [33] has the metric ansatz
\begin{equation}
%1
\label{eq1}
{\rm d}s^2=-f(r){\rm d}t^2+\frac{{\rm d}r^2}{f(r)}+r^2{\rm d}\varOmega _3^2 ,
\end{equation}
in which the $S^3$ line element can be expressed in the alternative form
\begin{equation}
%2
\label{eq2}
{\rm d}\varOmega_3^2 =\frac{1}{4}({\rm d}\theta ^2+{\rm d}\phi ^2+{\rm d}\psi ^2-2\cos \theta
{\rm d}\phi {\rm d}\psi ),
\end{equation}
where
\[
0\leqslant \theta \leqslant \uppi ,
\quad
0\leqslant \phi ,
\quad
\psi \leqslant 2\uppi .
\]
The energy-momentum tensor
\begin{equation}
%3
\label{eq3}
T_{\mu \nu } =2F_\mu ^{{\rm i}\alpha } F_{\nu \alpha }^{\rm i} -\frac{1}{2}g_{\mu \nu }
F_{\alpha \beta }^{\rm i} F^{{\rm i}\alpha \beta },
\end{equation}
where $F_{\alpha \beta }^{\rm i}$ is the Yang-Mills field 2-forms such
that $F_{\alpha \beta }^{\rm i} F^{{\rm i}\alpha \beta }={6Q^2}/{r^4}$, $Q$ is
the only non-zero gauge charge. The modified Einstein equations in EYMGB theory are [33]
\begin{eqnarray*}
&&G_{\mu \nu } -\alpha\Bigg[\frac{1}{2}g_{\mu \nu } (R_{\kappa \lambda \rho \sigma
} R^{\kappa \lambda \rho \sigma }-4R_{\rho \sigma } R^{\rho \sigma}+R^2)-2RR_{\mu \nu }\nonumber\\
&&+4R_\mu ^\lambda R_\nu ^\lambda +4R^{\rho \sigma }R_{\mu \rho \nu \sigma }
-R_\mu ^{\rho \sigma \lambda } R_{\nu \rho \sigma \lambda }\Bigg]=T_{\mu \nu }.
\end{eqnarray*}
The solution of field equations is given by
\begin{equation}
%4
\label{eq4}
f(r)=1+\frac{r^2}{4\alpha }\pm \sqrt {\left(\frac{r^2}{4\alpha}\right)^2+
\left(1+\frac{M}{2\alpha }\right)+\frac{Q^2\ln r}{\alpha }} ,
\end{equation}
in which $M$ is the usual integration constant to be identified as mass. In
the limit $\alpha \to 0$, the Einstein-Yang-Mills solution was obtained
with $M$ as a mass of the system, provided {the} negative branch of the above
solution is chosen as the form [22, 33]
\begin{equation}
%5
\label{eq5}
f(r)\to 1-\frac{M}{r^2}-\frac{2Q^2\ln r}{r^2}.
\end{equation}
The metric coefficient $f(r)$ in equation (\ref{eq3}) is identical to that of the
Einstein-Yang-Mills solution, and hence ``$M$" is interpreted as the mass of
the system. In {Eq.}~(\ref{eq4}) the expression within the square root is
positive definite for $\alpha >0$, {while} the geometry has a curvature
singularity at the surface $r=r_{\min }$ for $\alpha <0$. Here, $r_{\min}$
is the minimum value of the radial coordinate such that the function
under the square root is positive. Moreover, according to the values of the
parameters $(M,Q,\alpha )$, the singular surface can be surrounded by an
event horizon with radius $r_{\rm h}$. However, if no event horizon exists, there
will be a naked singularity.

Now the metric described by equation (\ref{eq4}) has a singularity at the greatest
real and positive solution ($r_{\rm s})$ of the equation
\begin{equation}
%6
\label{eq6}
\frac{r^4}{16\alpha ^2}+\left(1+\frac{M}{2\alpha }\right)+\frac{Q^2\ln r}{\alpha }=0.
\end{equation}
Note that if Eq.~(\ref{eq4}) has no real positive solution, then the metric
diverges at $r=0$. However, the singularity is surrounded by the event
horizon $r_{\rm h}$, which is the positive root of (the larger one if there are
two positive real roots)
\begin{equation}
%7
\label{eq7}
r^2-M-2Q^2\ln r=0.
\end{equation}
If $r_{\rm s} <r_{\rm h} $, {then} the singularity will be covered by the event horizon, while
the singularity will be naked for $r_{\rm s} \geqslant r_{\rm h}$. In this connection one may
note that the event horizon is independent of the coupling parameter $\alpha$.

As the event horizon $r_{\rm h}$ satisfies Eq.~(\ref{eq7}), {we} have
\begin{equation}
%8
\label{eq8}
M=r_{\rm h}^2 -2Q^2\ln r_{\rm h}.
\end{equation}
The surface area of the event horizon is given by $A=2\uppi ^2r_{\rm h}^3$. The
entropy of the black hole takes the form as
\begin{equation}
%9
\label{eq9}
S=\frac{c^3K_{\rm B} A}{4G\hbar }=\frac{c^3K_{\rm B} \uppi ^2}{2G\hbar }r_{\rm h}^3 .
\end{equation}
Considering $c=\hbar =G=1$, and the Boltzmann constant appropriately [22],
Eq.~(\ref{eq9}) can be expressed as
\begin{equation}
%10
\label{eq10}
S=r_{\rm h}^3 .
\end{equation}
From Eq.~(\ref{eq8}), $M$can be obtained as a function of $S$ and $Q$ in the form
\begin{equation}
%11
\label{eq11}
M=S^{2/3}-\frac{2}{3}Q^2\ln S.
\end{equation}
From the energy conservation law of the black hole ${\rm d}M=T{\rm d}S+\phi{\rm d}Q$, the
thermodynamic temperature and electric potential of the black hole can be given by
\begin{equation}
%12
\label{eq12}
T=\left(\frac{\vpartial M}{\vpartial S}\right)_Q
=\frac{2}{3}S^{-\frac{1}{3}}-\frac{2Q^2}{3S},
\end{equation}
and
\begin{equation}
%13
\label{eq13}
\phi =\left(\frac{\vpartial M}{\vpartial Q}\right)=-4Q\ln S.
\end{equation}
Now we turn to using the recent geometric formulation of {the} extended
thermodynamic behavior of the 5D black hole.

The formulation of {the} GTD of {the} black hole is based on the use of contact geometry
as a framework for thermodynamics [23]. Consider the (2$n$+1)-dimensional
thermodynamic phase space $\Im $ coordinates $Z^A=\{\phi ,E^a,I^a\}$ with
$A=0,\cdots ,2n$ and $a=1,\cdots,n$. In ordinary thermodynamics,
$\phi$ corresponds to {the} thermodynamic potential, and $E^a$ and $I^a$ are the
extensive and intensive variables, respectively. The fundamental
differential form $\varTheta $ can then be written in a canonical manner
as $\varTheta ={\rm d}\phi -\delta _{ab} I^a{\rm d}E^b$, where $\delta _{ab}$ is the
Euclidean metric. Consider $\Im $ {as} a non-degenerate metric $G=G(Z^A)$, and
the Gibbs1-form with $\delta _{ab} =\rm diag(1,1,\cdots,1)$. The set
$(\Im ,\varTheta,G)$ defines a contact Riemannian manifold if the condition $\varTheta \wedge
({\rm d}\varTheta )^n\ne 0$ is satisfied. This arbitrariness is restricted by the
condition that $G$ must be invariant with respect to Legendre transformations.
This is a necessary condition for our description of thermodynamic systems
{being} independent of the thermodynamic potential, {and} implies that
$\Im$ must be a curved manifold [23] because the special case of a metric with
vanishing curvature turns out to be {non-Legendre} invariant. The Gibbs 1-form
$\varTheta $ is also invariant with respect to Legendre transformations.
Legendre invariance guarantees that the geometric properties of $G$ do not
depend on the thermodynamic potential used in its construction.

The thermodynamic phase space {is} $\Im $, which in the case of the
(2+1)-dimensional black hole with a coulomb-like field, can be defined as a
{four}-dimensional space with coordinates $Z^A=\{M,S,T,Q\}$, $A=0,\cdots,3$.
Eq.~(\ref{eq11}) represents the fundamental relationship $M=(S,Q)$ from which
all {thermodynamic} information can be obtained, therefore we would like
to consider a {four}-dimensional phase space $\Im $ with coordinates $(M,S,T,Q)$,
a contact1-form
\begin{equation}
%14
\label{eq14}
\varTheta ={\rm d}M-T{\rm d}S-\phi{\rm d}Q,
\end{equation}
and an invariant metric
\begin{equation}
%15
\label{eq15}
G=({\rm d}M-T{\rm d}S-\phi{\rm d}Q)^2+(TS+\phi Q)(-{\rm d}T{\rm d}S+{\rm d}\phi{\rm d}Q).
\end{equation}
The triplet $(\Im ,\varTheta ,G)$ defines a contact Riemannian manifold that
plays an auxiliary role in GTD. It is used to properly handle the invariance
with respect to Legendre transformations. In fact, for the charged black
hole, a Legendre transformation involves in general all the thermodynamic
variables $M$, $S$, $Q$, $T$ and $\phi $, so that they must be independent
of each other as they are in the phase space. We also introduce the
geometric structure of the space of equilibrium states $\varepsilon $ in the
following manner: $\varepsilon $ is a {two}-dimensional submanifold of $\Im $
that is defined by the smooth embedding map $\varphi $: $\varepsilon \mapsto
\Im $, satisfying the condition that the ``projection'' of the contact form
$\varTheta $ on $\varepsilon $ vanishes, namely $\varphi ^\ast (\varTheta )=0$, where
$\varphi ^\ast $ is the pullback of $\varphi $, and that $G$ induces a Legendre
invariant metric $g$ on $\varepsilon $ by means of $\varepsilon $. In
principle, any {two}-dimensional subset of the set of coordinates of $\Im $ can
be used to coordinative $\varepsilon $. For the sake of simplicity, we will
use the set of extensive variables $S$ and $Q$, which corresponds to the
energy representation in ordinary thermodynamics. Then, the embedding map
for this specific choice is $\varphi $: $\{S,Q\}\mapsto
\{M(S,Q),S,Q,T(S,Q),\phi (S,Q)\}$. The condition $\varphi ^\ast (\varTheta )=0$
is equivalent to the first law of thermodynamics and the conditions of
thermodynamic equilibrium
\begin{equation}
%16
{\rm d}M=T{\rm d}S+\phi{\rm d}Q,~T=\frac{\vpartial M}{\vpartial S},~\phi
=\frac{\vpartial M}{\vpartial Q},
\end{equation}
whereas the induced metric becomes
\begin{equation}
%17
\label{eq16}
g=\left(S\frac{\vpartial M}{\vpartial S}+Q\frac{\vpartial M}{\vpartial
Q}\right)\left(-\frac{\vpartial ^2M}{\vpartial S^2}{\rm d}S^2+\frac{\partial ^2M}{\vpartial
Q^2}{\rm d}Q^2\right).
\end{equation}

This metric determines all the geometric properties of the equilibrium
space $\varepsilon $. In order to obtain the explicit form of the metric, it
is necessary to specify the thermodynamic potential $M$ as a function of
$S$ and $Q$. In ordinary thermodynamics this function is usually referred to
as the fundamental equation from which all the equations of state can be derived.

Substituting Eq.~(\ref{eq11}) into Eq.~(\ref{eq16}), we can obtain the Legendre metric
components of the 5D black hole as
\begin{eqnarray}
%18
\label{eq17}
g&=&\frac{4\left( {3Q^2-S^{2/3}} \right)\left( {Q^2-S^{2/3}+2Q^2\ln S}
\right)}{27S^2}{\rm d}S^2\nonumber\\[3mm]
&&+\frac{8}{9}\left( {Q^2-S^{2/3}+2Q^2\ln S} \right)\ln S{\rm d}Q^2.
\end{eqnarray}
After some calculations, the Legendre metric scalar curvature is obtained
\begin{eqnarray}
%19
\label{eq18}
\Re =\frac{N}{8(\ln S)^2(S^{2/3}-3Q^2)^2(Q^2\ln S-S^{2/3}+Q^2)^3}.
\end{eqnarray}
\begin{eqnarray}
%20
\label{eq19}
{N}&=&27(Q^2-S^{2/3})(3Q^2-S^{2/3})+18(9Q^6+8Q^4S^{2/3}\nonumber\\[3mm]
&&-13Q^2S^{4/3}+4S^2)\ln S+24(27Q^6+3Q^4S^{2/3}\nonumber\\[3mm]
&&-11Q^2S^{4/3}+S^2)(\ln S)^2+48Q^2S^{4/3}(\ln S)^3.
\end{eqnarray}

Obviously, the non-flatness of the Legendre metric indicates that the black
hole thermodynamics {have} statistical mechanical interactions. According to
Davies' approach [34], the phase transition structure of the 5D black hole
can be derived from the heat capacity
\begin{equation}
%21
\label{eq20}
C_Q =T\left(\frac{\vpartial S}{\vpartial T}\right)_Q =\frac{3S(S^{2/3}-Q^2)}
{3Q^2-S^{2/3}}.
\end{equation}
Hence, $C_Q =0$ at $S^{2/3}=Q^2$. Another interesting point is
\begin{equation}
%22
\label{eq21}
3Q^2=S^{2/3}.
\end{equation}

At this point, $C_Q $ changes sign and the scalar curvature {diverges}.
Therefore, there is a phase transition which corresponds to this critical point. This behavior is shown in Fig.~1.

%fig 1
\begin{center}
\includegraphics{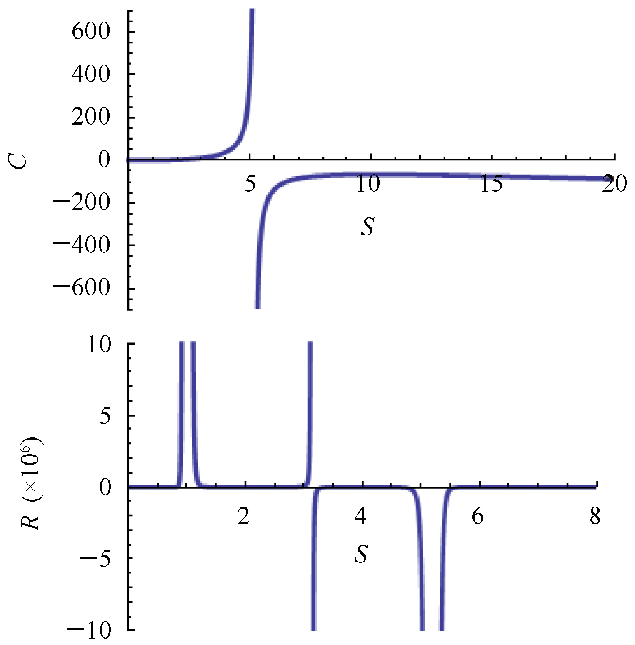}
\figcaption{{The} behavior of the heat capacity and scalar curvature as functions of
the entropy of the black hole with $Q=1$. {Their} singularities are located at $S\approx 5$.}
\end{center}

\section{Geometrothermodynamics of the 5D black hole in EMGB theory}

The action of the {five-dimensional} space time $(M$, $g_{\mu \nu })$ that
represents Einstein-Maxwell-Gauss-Bonnet theory with a cosmological constant
is {given} by [22, 35, 36]
\begin{equation}
%23
\label{eq22}
S=\frac{1}{2}\vint\nolimits_M {{\rm d}^5x} \sqrt {-g}\left [R-2\varLambda -\frac{1}{4}F_{\mu \nu }
F^{\mu \nu }\alpha R_{GB}\right ],
\end{equation}
where
\begin{equation}
%24
\label{eq23}
R_{\rm GB} =R^2-4R_{\alpha \beta } R^{\alpha \beta }+R_{\alpha \beta \gamma
\delta } R^{\alpha \beta \gamma \delta },
\end{equation}
which is the Gauss-Bonnet term. $\alpha $ is the GB coupling parameter {with a}
dimension (length)$^2$, $\varLambda $ is the cosmological constant and $F_{\mu
\nu } =(\vpartial _\mu A_\nu -\vpartial _\nu A_\mu )$ is the usual
electromagnetic field tensor with $A_\mu $, the vector potential.

The {five-dimensional} spherically symmetric solution obtained has the line element
\begin{equation}
%25
\label{eq24}
{\rm d}s^2=-B(r){\rm d}t^2+B^{-1}(r){\rm d}r^2+r^2({\rm d}\theta _1^2 +\sin ^2\theta _1 ({\rm d}\theta
_2^2 +\sin ^2\theta _2{\rm d}\theta _3^2 )),
\end{equation}
with
$$0\leqslant \theta _1 ,~ \theta _2\leqslant\uppi ,~ 0\leqslant \theta _3 \leqslant 2\uppi .$$

Solving the above field equations, one obtains [22, 34]
\begin{equation}
%26
\label{eq25}
B(r)=1+\frac{r^2}{4\alpha }-\frac{r^2}{4\alpha }\sqrt {1+\frac{16M\alpha
}{\uppi r^4}-\frac{8Q^2\alpha }{3r^6}+\frac{4\varLambda \alpha }{3}} .
\end{equation}
In an orthonormal frame, the non-null components of the electromagnetic
tensors $F_{\hat {t}\hat {r}} =-F_{\hat {r}\hat {t}} =Q/{4\uppi r^3}$.
Ref.~[22] {gives} an equation about Eq.~(\ref{eq25}), is well defined for
$r>r_{\min}$ and $r_{\min }$ satisfies
\begin{equation}
%27
\label{eq26}
1+\frac{16M\alpha }{\uppi r_{\min }^4 }-\frac{8Q^2\alpha }{3r_{\min }^6
}+\frac{4\varLambda \alpha }{3}=0.
\end{equation}
The surface $r=r_{\min } $ corresponds to a curvature singularity. However,
depending on the values of the parameters, this singular surface may be
surrounded by the event horizon ($B(r_{\rm h})=0)$, and the solution (\ref{eq26})
describes a black hole solution known as the EMGB black hole.

The metric depends on two parameters, $Q$ and $M$, which are identified with
the electric charge and the mass, respectively.
\begin{equation}
%28
\label{eq27}
M=\uppi \alpha +\frac{\uppi Q^2}{6}r_{\rm h}^{-2} +\frac{\uppi r_{\rm h}^2 }{2}-\frac{\uppi
\varLambda }{12}r_{\rm h}^4 .
\end{equation}
From Eq.~(\ref{eq10}), Eq.~(\ref{eq30}) can be expressed as
\begin{equation}
%29
\label{eq28}
M=\uppi \alpha +\frac{\uppi Q^2}{6}S^{-2/3}+\frac{\uppi S^{2/3}}{2}-\frac{\uppi \varLambda }{12}S^{4/3}.
\end{equation}
This equation relates all the thermodynamic variables entering the EMGB
black hole so that if we impose the first law of
thermodynamics ${\rm d}M=T{\rm d}S+\phi{\rm d}Q$, the expressions for the temperature and
the electric potential can be easily computed as $T={\vpartial M}/{\vpartial S},~\phi =
{\vpartial M}/{\vpartial Q} $. It is convenient to write the
final results in terms of the horizon radii by using the relations
\begin{equation}
%30
\label{eq29}
T=\left(\frac{\vpartial M}{\vpartial S}\right)_Q =\frac{\uppi}{3}S^{-1/3}-\frac{\uppi}
{9}Q^2S^{5/3}-\frac{\uppi \varLambda }{9}S^{1/3},
\end{equation}
and
\begin{equation}
%31
\label{eq30}
\phi =\left(\frac{\vpartial M}{\vpartial Q}\right)=\frac{\uppi Q}{3}S^{{-2}/3},
\end{equation}
Substituting Eq.~(\ref{eq28}) into Eq.~(\ref{eq16}), we can obtain the Legendre metric
components of the EMGB black hole as
\end{multicols}
\begin{eqnarray}
%32
\label{eq31}
g&=&-\frac{\left( {-2\uppi Q^2-3\uppi S^{4/3}+\uppi \varLambda S^2} \right)
\left({-5\uppi Q^2+3\uppi S^{4/3}+\uppi \varLambda S^2} \right)}{243S^{10/3}}{\rm d}S^2
+\frac{\uppi \left( {2\uppi Q^2+3\uppi S^{4/3}-\uppi \varLambda S^2}
\right)}{27S^{4/3}}{\rm d}Q^2.
\end{eqnarray}
\vspace{2mm}
\begin{multicols}{2}
\noindent After some calculations, the Legendre invariant scalar curvature can be expressed as
\begin{eqnarray}
%33,34
\label{eq32}
\Re &=& \frac{54S^{8/3}N}{(2\uppi Q^2+3\uppi S^{4/3}-\uppi \varLambda
S^2)(-5\uppi Q^2+3\uppi S^{4/3}+\uppi \varLambda S^2)},\nonumber\\ \\[3mm]
\label{eq33}
N &=& 27\uppi ^3(4Q^4-10Q^2S^{4/3}-3S^{8/3})+\uppi ^3
\varLambda S^{2/3}(104Q^4\nonumber\\[3mm]
&&-30Q^2S^{4/3}+63S^{8/3})-10\uppi \varLambda ^3S^{{14}/3}\nonumber\\[3mm]
&&+\uppi \varLambda ^3(4Q^2S^{8/3}+6S^2).
\end{eqnarray}
Obviously, the non-flatness of the Legendre metric indicates that the black
hole thermodynamics {have} statistical mechanical interactions. The expression
for the heat capacity with a fixed charge is given by
\begin{equation}
%35
\label{eq34}
C_Q =\frac{3S\left( {\uppi Q^2-3\uppi S^{4/3}+\uppi \varLambda S^2} \right)}
{-5\uppi Q^2+3\uppi S^{4/3}+\uppi \varLambda S^2}.
\end{equation}

The heat capacity has a zero-point at $Q^2=3S^{4/3}-\varLambda S^2$.

%fig 2
\begin{center}
\includegraphics{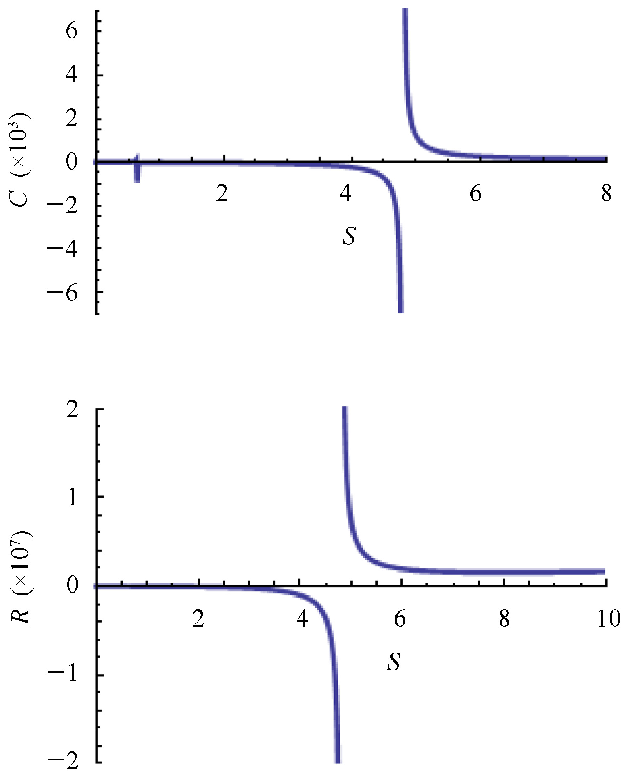}
\figcaption{{The} behavior of the heat capacity and scalar curvature as functions of
the entropy of the black hole with $Q=1/2$. {Their} singularities are located at $S\approx 4.9$.}
\end{center}

\noindent Moreover,
the sign change of the heat capacity and the divergence of the scalar
curvature occur at $5Q^2=3S^{4/3}+\varLambda S^2$. Therefore, there will be a
phase transition at this point. This behavior is shown in Fig.~2.

\section{Discussions and conclusions}

In this work, we reproduced {thermodynamic} properties such as {the} temperature
and entropy of the 5D black hole in Gauss-Bonnet gravity theory. We studied
the Legendre invariant metric of the 5D black hole. The results show that
GTD delivers a particular thermodynamic metric for the 5D black hole and the
EMGB black hole. Then we {corroborated} that the thermodynamic curvature is
non-zero and its singularities reproduce the phase transition structure, which
follows from the divergences of the heat capacity.

In addition, the thermodynamic metric proposed in this work {was} applied
to the case of black hole configurations in three dimensions. It {was}
shown that this thermodynamic metric correctly describes the thermodynamic
behavior of the corresponding black hole configurations. One additional
advantage of this thermodynamic metric is its invariance with respect to
total Legendre transformations. This means that the results are independent
of the thermodynamic potential used to generate the thermodynamic metric. In
all the remaining cases, the singularities of the thermodynamic curvature
correspond to points where the heat capacity diverges and phase transitions
takes place. We interpret this result as an additional indication that the
thermodynamic curvature, as defined in GTD, can be used to measure {the}
thermodynamic interaction. In fact, it has been shown that in the case of
more realistic thermodynamic systems [27], the ideal gas is also
characterized by a vanishing thermodynamic curvature, whereas the van der
Waals gas generates a nonvanishing curvature whose singularities {reproduce}
the corresponding phase transition structure.

Furthermore, we expect that this unified geometry description may give more
information about a thermodynamic system. We also conclude that GTD is, in
general, duality invariant. Therefore, our results support Quevedo's viewpoint.

\end{multicols}

\vspace{-2mm}
%\centerline{\rule{80mm}{0.1pt}}
\vspace{2mm}

\begin{multicols}{2}

%参考文献

\end{multicols}

\clearpage

\end{document}